\documentclass[reprint,amsmath,amssymb,aps,pre]{revtex4-1}
\usepackage[dvips]{graphicx}
\usepackage{bm}
\usepackage{braket}
\usepackage{mathtools}

\begin{document}
\title{
  Measurement of entanglement entropy in the two-dimensional Potts model using wavelet analysis
}
\author{Yusuke Tomita}
\affiliation{College of Engineering, Shibaura Institute of Technology, Saitama, Saitama 337-8570, Japan}

\date{\today}

\begin{abstract}
  We introduce a method to measure the entanglement entropy using
  a wavelet analysis.
  In the method we perform the two-dimensional Haar wavelet transform
  of configuration of Fortuin-Kasteleyn (FK) clusters.
  The configuration represents a direct snapshot of spin-spin correlations
  since spin degrees of freedom are traced out in FK representation.
  A snapshot of FK clusters loses image information
  at each coarse-graining process by the wavelet transform.
  We show that the loss of image information measures
  the entanglement entropy in the Potts model.
\end{abstract}

\maketitle

\section{%
  \label{sec:intro}
  Introduction
}

To represent a many-body system using minimal bases
has been a challenging subject in numerical renormalization-group (RG)
calculations~\cite{White1992,White1993,NishinoOkunishi1996,NishinoOkunishi1997,LevinNave2007,EvenblyVidal2009,EvenblyVidal2015}.
Developing numerical RG methods not only enables us
to calculate larger many-body systems in a shorter time,
but also gives us deeper insights into many-body physics
because representing a many-body system using minimal bases
is equivalent to extracting the essence of the system.

In the 1970s and 1980s, combinational use of the Monte Carlo simulation
and the real-space RG~\cite{Ma1976,Swendsen1979}
gave a powerful tool to investigate critical phenomena.
However, due to difficulties to erase or suppress side defects
caused by the block spin transformation,
the Monte Carlo RG approach had been stagnated.
Recently, the idea of typicality of thermal equilibrium~\cite{SugiuraShimizu2013,Tasaki2016}
motivates to utilizing a snapshot of spin configuration as a tool
to investigate many-body systems.
To the best of the author's knowledge, Ueda and his collaborators
are the first who studied a relation between a snapshot
of spin configuration and the variational state employed in
the corner transfer matrix RG~\cite{Ueda2005}.
Matsueda proposed a more direct method to investigate a snapshot
of a spin configuration~\cite{Matsueda2012}.
He found that the singular value decomposition (SVD) of a snapshot
reveals a hierarchical structure in spin configuration patterns in a snapshot.
He also defined a snapshot entropy and discussed
the truncation number of the SVD dependence of the quantity.
Finite-size effect of the snapshot entropy, however,
is not clear even though several studies have been conducted~\cite{Matsueda2012,Imura2014,Lee2015,MatsuedaOzaki2015,Lee2016}.
Poor understanding of the snapshot entropy has hampered to
establish a method to study many-body systems using a snapshot of a spin configuration.

In this paper, we propose a method that provides a means to measure
the entanglement entropy using
a wavelet analysis~\cite{Daubechies}.
An important modification to the preceding method
is that we do not focus on a spin configuration but
on a Fortuin-Kasteleyn (FK) cluster configuration~\cite{KasteleynFortuin1969,FortuinKasteleyn1972}.
Our results indicate that a mere spin configuration is too superficial
to extract an essence of a many-body system.
An entropy given in this paper has an advantage that
it follows Calabrese-Cardy formula~\cite{Holzhey1994,CalabreseCardy2004}.

This paper is organized as follows.
In Sec.~\ref{sec:model}, we give the Hamiltonian of the Potts model
and its percolation formula.
A wavelet analysis used in this paper is explained by using a one-dimensional (1D) model
in Sec.~\ref{sec:method}.
Numerical results are given in Sec.~\ref{sec:results}.
Section~\ref{sec:discussion} is devoted to
the summary and discussion.

\section{%
  \label{sec:model}
  Model
}

In this paper, we apply the two-dimensional (2D) Haar wavelet transform to
the 2D ferromagnetic $q$-state Potts model on the square lattice.
The Hamiltonian of the $q$-state Potts model~\cite{Potts1952,Kihara1954,Wu1982} is given by
\begin{equation}
  \label{eq:hamiltonian}
  \mathcal{H} = -J\sum_{\langle i,j\rangle}\delta_{\sigma_i,\sigma_j},
\end{equation}
where $J$ is the exchange coupling constant,
$\sigma_i$ is a spin at site $i$,
and $\delta_{\sigma_i,\sigma_j}$ is the Kronecker delta.
The partition function of the Potts model $Z$ can be written
in the percolation formula~\cite{KasteleynFortuin1969,FortuinKasteleyn1972,ConiglioKlein1980,Hu1984a},
\begin{equation}
  \label{eq:percolation_formula}
  Z = e^{N_b\beta J}\sum_{g\subseteq G}p^{b(g)}(1-p)^{N_b-b(g)}q^{c(g)},
\end{equation}
where $\beta$ is the inverse temperature,
$N_b$ is the total number of bonds,
$G$ is all the bond configurations,
$g$ is a bond configuration,
$p$ is the probability of connecting the same spins,
$b(g)$ is a number of occupied bonds in a bond configuration $g$,
and $c(g)$ is a number of clusters in $g$.
The probability $p$ depends on the temperature,
\begin{equation}
  \label{eq:probability}
  p = 1 - e^{-\beta J}.
\end{equation}
Hereafter, Boltzmann constant $k_{\mathrm{B}}$ and
the exchange coupling constant $J$ are set to unity.

\section{%
  \label{sec:method}
  Method
}

The quantity we focus on in the paper is a sum of squared wavelet
coefficients.
I define the quantity $\mathcal{E}$, entropy emission, as
\begin{equation}
  \label{eq:entropy_emission}
  \mathcal{E}_m = 4^{m+1}\sum_{i,j}[(d^v_{m;i,j})^2 + (d^h_{m;i,j})^2 + (d^d_{m;i,j})^2],
\end{equation}
where $m$ is the number of level, and $d^\alpha$'s are wavelet coefficients (see Appendix).
Using $\epsilon_m$ we denote the normalized entropy emission,
\begin{equation}
  \label{eq:entropy_emission_ps}
  \epsilon_m = \frac{4^{m+1}}{3N_m}
  \sum_{i,j} [(d^v_{m;i,j})^2 + (d^h_{m;i,j})^2 + (d^d_{m;i,j})^2],
\end{equation}
where $N_m(=2^{2(n-m)})$ is the number of level-$m$ sites.
To see properties of the quantity, we exemplify a simple one-dimensional model.
A function $f(x)$ is defined on the one-dimensional
lattice sites of length $L(=2^n)$.
Using level-$0$ scaling function $\phi_{0;i}(x)$,
$f(x)$ is represented by
\begin{equation}
  f(x) = \sum_{i}c_{0;i}\phi_{0;i}(x),
\end{equation}
where $c_{0;i}$ is a scaling coefficient.
The function $\phi_{0;i}(x)$ is defined by
\begin{equation}
  \label{eq:haar_scaling1d}
  \phi_{0;i}(x) = \left\{
  \begin{array}{ll}
    1 & (\mbox{at $x = i$}),\\
    0 & (\mbox{otherwise}).
  \end{array}
  \right.
\end{equation}
Scaling functions of higher level are recursively defined by
\begin{equation}
  \label{eq:phi_recursion_1d}
  \phi_{m;i}(x) = \phi_{m-1;2i}(x) + \phi_{m-1;2i+1}(x).
\end{equation}
Using level-$m$ scaling functions,
the coarse-grained function $f_m(x)$ is represented by
\begin{align}
  \label{eq:f_level_m_1}
  f_m(x) &= \sum_{i}\braket{f(x)|\phi_{m;i}(x)}\phi_{m;i}(x),\\
  \label{eq:f_level_m_2}
  &= \sum_{i}c_{m;i}\phi_{m;i}(x),
\end{align}
where $\braket{f(x)|g(x)}$ denotes the inner product,
$\sum_xf(x)g(x)$.
Note that the scaling functions are orthogonal;
\begin{equation}
  \label{eq:orthogonarity1d}
  \braket{\phi_{m;i}(x)|\phi_{m;i'}(x)} = 2^m\delta_{i,i'}.
\end{equation}
The recursion relation of the scaling functions,
Eq.~(\ref{eq:phi_recursion_1d}),
deduces the following relation for neighboring level coefficients,
\begin{equation}
  c_{m;i} = \frac{1}{2}(c_{m-1;2i} + c_{m-1;2i+1}).
\end{equation}
That is, coarse-graining procedure using the Haar wavelet
is equivalent to an averaging procedure.

Wavelet coefficients preserve
lost information caused by the coarse-graining.
The 1D Haar wavelet functions of level-$m$ are given by
the difference of neighboring scaling function of level-$(m-1)$,
\begin{equation}
  \label{eq:haar_wavelet1d}
  \psi_{m;i}(x) = \frac{1}{2^m}[\phi_{m-1;2i}(x)-\phi_{m-1;2i+1}(x)].
\end{equation}
The wavelet functions succeed the orthogonality of
the scaling functions,
\begin{equation}
  \braket{\psi_{m;i}(x)|\psi_{m;i'}(x)} = \frac{1}{2^m}\delta_{i,i'}.
\end{equation}
The wavelet coefficients are obtained by the following inner product,
\begin{equation}
  d_{m;i} = \braket{f(x)|\psi_{m;i}(x)}.
\end{equation}
As in the coefficients of scaling functions,
the wavelet coefficients have relation for neighboring level coefficients,
\begin{equation}
  d_{m;i} = \frac{1}{2}(c_{m-1;2i} - c_{m-1;2i+1}).
\end{equation}
From the relation, we can see that wavelet coefficients
preserve the function form of $f_{m-1}(x)$ as differences of
scaling function coefficients of level-$(m-1)$.
In the following demonstration,
we modify the definition of the entropy emission as,
\begin{equation}
  \label{eq:entropy_emission_1d}
  \mathcal{E}_m = 2^{m+2}\sum_{i} (d_{m;i})^2,
\end{equation}
since we consider only one type of wavelet function.

\begin{figure*}[ht]
  \includegraphics[width=0.85\textwidth]{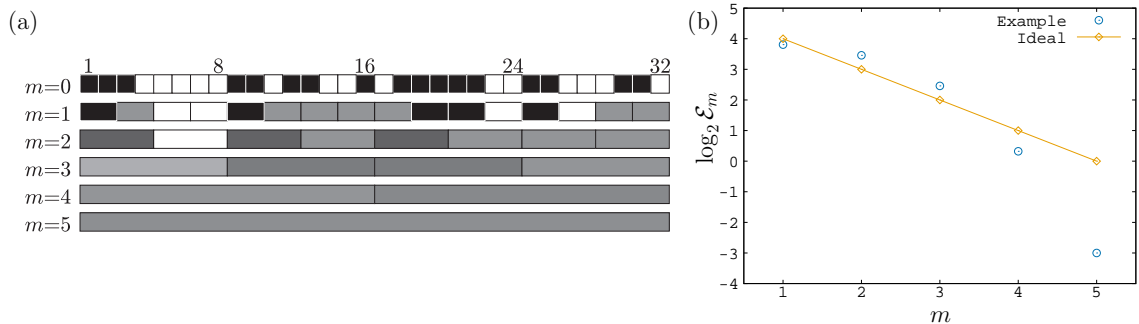}
  \caption{\label{fig:ex1d}
    A demonstration of the 1D Haar wavelet transform.
    (a) Coarse-graining steps for each level-$m$ are shown.
    Black (white) squares at $m=0$ denote that the correspondent
    bits are zero (unity).
    An intensity of a gray rectangle denotes its coarse-grained value.
    (b) Plot of the logarithm of the entropy emission demonstrated in Fig.~\ref{fig:ex1d}(a).
    Orange line and symbols denote an \textit{ideal} dependence on $m$,
    $\log_2\mathcal{E}_m = \log_2 L-m$.
  }
\end{figure*}

Using the 1D Haar wavelet, we demonstrate the wavelet transformation
of a binary random number sequence of length 32.
The original random number sequence, $f(x)$,
is shown in the top row of Fig.~\ref{fig:ex1d}(a);
black and white squares respectively denote 0 and 1.
The entropy of 32-bit random number is 32 if we take the logarithm base 2.
Intensities of gray rectangles in Fig.~\ref{fig:ex1d}(a)
denote values of the coarse-grained functions; by definition, the range
of values is 0 to 1.
In this demonstration, entropy emissions
given in Eq.~(\ref{eq:entropy_emission_1d})
are $\mathcal{E}_1 = 14$,
$\mathcal{E}_2 = 11$,
$\mathcal{E}_3 = 5.5$,
$\mathcal{E}_4 = 1.25$,
and $\mathcal{E}_5 = 0.125$.
The accumulated entropy emission is 31.875,
which is close to 32, the entropy of 32-bit random number sequence.
Actually, the value of accumulated entropy emission comes closer
to the bit length as the length is increased.
From the demonstration we see that the quantity $\mathcal{E}_m$ shows
the amount of entropy emission in a coarse-graining procedure between
$(m-1)$th and $m$th wavelet transformation.

\section{%
  \label{sec:results}
  Results
}

\begin{figure*}[ht]
  \includegraphics[width=0.85\textwidth]{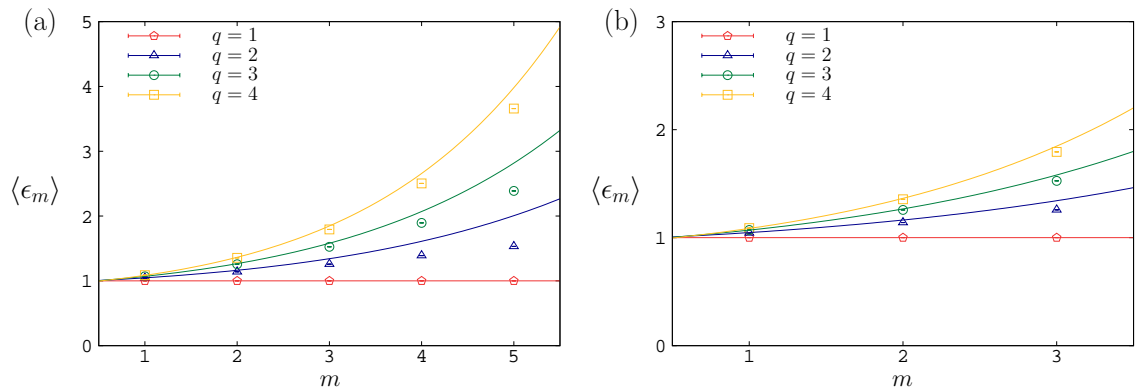}
  \caption{\label{fig:entropy_emission_2d}
    (a) $m$ dependence of normalized entropy emissions
    of the 2D $q$-state Potts models.
    Curves denote Calabrese-Cardy formula, Eq.~(\ref{eq:calabrese_cardy2}).
    (b) Magnified view of Fig.~\ref{fig:entropy_emission_2d}(a).
    Error bars are smaller than the size of the symbols.
  }
\end{figure*}

Entropy emissions of the 2D $q$-state Potts models at critical points are
calculated by Monte Carlo simulation.
System size $L$ is $16384(=2^{14})$.
A number of samples is $10^4$, and 10 independent Monte Carlo
runs are executed to estimate statistical errors.
In order to equilibrate spin configurations for $q\ge 2$ models,
Swendsen-Wang algorithm~\cite{SwendsenWang1987, KomuraOkabe2012, Komura2015}
is used, and
$2\times 10^4$ Monte Carlo steps are discarded for thermalization.
The temperature is fixed at the critical point of the 2D $q$-state Potts model,
\begin{equation}
  \label{eq:critical_temp}
  T_{\mathrm{c}} = \frac{1}{\ln(1 + \sqrt{q})}.
\end{equation}
In the percolation formula, this gives the critical probability of
connecting the same spins [see Eq.~(\ref{eq:percolation_formula})],
\begin{equation}
  \label{eq:critical_bond}
  p_{\mathrm{c}} = \frac{\sqrt{q}}{1+\sqrt{q}}.
\end{equation}
We execute the wavelet transform for each bond configuration
obtained during measurement runs.
The wavelet transform applies to each
$2\times 2$ block of vertical and horizontal bonds,
and the both of vertical and horizontal contributions
are taken into the entropy emission.
Figure~\ref{fig:entropy_emission_2d} shows
normalized entropy emissions for the 2D Potts models.
As we saw in the example of a random number sequence in Sec.~\ref{sec:method},
the quantity does not change its value
when site- or bond-variables are totally random.
Curves in Fig.~\ref{fig:entropy_emission_2d} are given by
\begin{equation}
  \label{eq:carabrese_cardy1}
  \epsilon_m = \epsilon_0\left(1 + \frac{c}{6}\mathcal{A}\ln s_m\right),
\end{equation}
where $c$ is the central charge, $\mathcal{A}$ is the boundary length
between clusters, and $s_m(=2^m)$ is a side length of a wavelet basis.
The second term in the parenthesis
in the right-hand side is known as
Calabrese-Cardy formula~\cite{Holzhey1994,CalabreseCardy2004}.
In the case of the 2D critical Potts model,
$\mathcal{A}$ is proportional to $s_m^{D_{\mathrm{EP}}-1}$,
where $D_{\mathrm{EP}}$ is the dimension of the external perimeter.
Exact values of $c$'s and $D_{\mathrm{EP}}$'s are listed in
Table~\ref{tab:c_and_dep}~\cite{Duplantier2000}.
To draw curves in Fig.~\ref{fig:entropy_emission_2d},
we assumed the proportional constant as unity; that is,
\begin{equation}
  \label{eq:calabrese_cardy2}
  \epsilon_m = \epsilon_0\left(1 + \frac{c}{6}2^{(D_{\mathrm{EP}}-1)m}\ln 2^m\right).
\end{equation}
Probably the value of the true proportional constant is close to unity,
since curves run on numerical data at $m=2$.

\begin{table}[h]
  \caption{%
    \label{tab:c_and_dep}
    Central charges $c$ and the dimensions of the external perimeter $D_{\mathrm{EP}}$ for the critical $q$-state Potts model~\cite{Duplantier2000}.
  }
  \begin{ruledtabular}
    \begin{tabular}{ccccc}
      $q$ & 1 & 2 & 3 & 4\\ \hline
      $c$ & 0 & 1/2 & 4/5 & 1\\
      $D_{\mathrm{EP}}$ & 4/3 & 11/8 & 17/12 & 3/2
    \end{tabular}
  \end{ruledtabular}
\end{table}

\section{%
  \label{sec:discussion}
  Summary and Discussion
}

The entropy emission seems to measure the entanglement entropy
in the 2D $q$-state Potts models, and it is curious how the entropy emission
can measure the entanglement entropy from a snapshot.
Wavelet coefficients detect segments of boundaries between FK clusters.
Level $m$ dependence of the coefficients has a correspondence with a characteristic length
of boundary:
Sizes of FK clusters are small at a high temperature,
and the entropy emission decreases as $m$ increases.
At a low temperature, to the contrary, sizes of FK clusters are large,
and the entropy emission increases as $m$ increases.
At an intermediate temperature, near a criticality, 
the entropy emission is almost flat.
An extreme case is the percolation problem at the critical point, where
the entropy emission is invariant under the wavelet transformation
due to the fractality of clusters' shape.
Entropy emissions of Potts models for $q=2,3,4$ at their criticalities
mildly increase as $m$ is increased.
Difference between the percolation problem and the spin models
is correlations between bonds.
The density of bonds at their criticality is universal, $b(g)/N_b=1/2$.
This comes from the self-dual of the square lattice.
Bond distributions of the spin models are more denser than that
of the percolation problem since bond occupation is restricted
between the same spins.
Dense FK clusters are robust against the wavelet transform,
and the robustness causes
the mild increase of the entropy emission.
Fractality of the FK clusters brings about non-trivial growth
of boundary length, $\mathcal{A} \propto s^{D_{\mathrm{EP}}-1}$,
and Eq.~(\ref{eq:calabrese_cardy2}) is deduced.
In the percolation formula, only a freedom of spin assignment
on each of FK clusters remains; that is, wavelet coefficients,
which are non-zero at a cluster boundary,
capture a universality class of a system from a snapshot
of a bond configuration at each level $m$.
To examine the conjecture formula, Eq.~(\ref{eq:calabrese_cardy2}),
elaborate studies on wavelet transformation of FK clusters are required.

The fact that spin configuration is not essential
but FK cluster configuration given by the percolation formula is important
suggests a guideline studying critical phenomena using image processing.
It has been known that the percolation formula is quite helpful
for a study of critical phenomena~\cite{TomitaOkabe2001,TomitaOkabeHu1999,SwendsenWang1987}, and
the analysis with wavelets confirms again its importance.
The insight promises that applications to world line configurations
in quantum spin Monte Carlo simulations~\cite{Evertz1993,KawashimaHarada2004,TomitaOkabe2002}
will work well.
I leave the application for a future study.
It is obvious that a wavelet transform of spin configuration
will fail to investigate critical phenomena of spin systems.
Since wavelet transform conserves magnetization,
its coarse-graining procedure deviates from the legitimate RG.
In this study, the wavelet transform is carried out for bond configuration,
and the densities of bonds are being kept correctly at 1/2
during the wavelet transform.
It is probable that
an inference of bond distribution is indispensable
when one investigates critical phenomena of spin systems
by image processing of spin configuration only.

Wavelet analysis proposed in this study is imperfect.
Except for the percolation problem,
deviations of normalized entropy emissions from Eq.~(\ref{eq:calabrese_cardy2})
systematically grows larger
with each wavelet transform.
The flaw would come from interfusion of bonds that
belongs to different FK clusters.
More sophisticated way to coarse-grain bonds will
restore the flaw, and the modification will bring about
clearer understanding on
relations between spin systems at criticality
and image processing.

\hfill

\begin{acknowledgments}
  The author thanks H. Matsueda for bringing his attention to
  the relation between snapshots and the entanglement entropy.
  This work was supported by JSPS KAKENHI Grant Number 16K05482.
\end{acknowledgments}

\begin{widetext}

\appendix*
\section{%
  \label{sec:appendix}
  Wavelet transform with the 2D Haar wavelet
}

In this paper, we used the 2D Haar wavelet transform
to obtain coarse-grained bond configuration.
Here, in stead of a function of bonds,
we consider a wavelet transform of a function defined
on a square lattice points of size $2^n\times 2^n$ for a brief description.
A level-$m$ coarse-grained bond configuration at a lattice point $(x,y)$,
$f_m(x,y)$,
is represented by a sum of level-$m$ scaling functions $\phi_{m;i,j}(x,y)$,
\begin{equation}
  \label{eq:coarse_grained_f}
  f_m(x,y) = \sum_{i,j}c_{m;i,j}\phi_{m;i,j}(x,y),
\end{equation}
where $c_{m;i,j}$ is a scaling coefficient.
Using level-$0$ scaling functions the original function $f(x,y)$
is represented by
\begin{equation}
  \label{eq:original_f}
  f(x,y) = \sum_{i,j}c_{0;i,j}\phi_{0;i,j}(x,y).
\end{equation}
The function $\phi_{0;i,j}(x,y)$ is defined by
\begin{equation}
  \label{eq:phi_0}
  \phi_{0;i,j}(x,y) =
  \left\{
  \begin{array}{ll}
    1 & \mbox{if $(x,y) = (i,j)$},\\
    0 & \mbox{otherwise}.
  \end{array}
  \right.
\end{equation}
Level-$m$ scaling functions can be expressed by
using level-$(m-1)$ scaling functions as
\begin{align}
  \label{eq:phi_m_1}
  \phi_{m;i,j}(x,y)
  &=
  \Braket{
  \left(
  \begin{array}{ll}
    1 & 1\\
    1 & 1
  \end{array}
  \right)|
  \left(
  \begin{array}{ll}
    \phi_{m-1;2i,2j}(x,y) & \phi_{m-1;2i+1,2j}(x,y)\\
    \phi_{m-1;2i,2j+1}(x,y) & \phi_{m-1;2i,2j+1}(x,y)
  \end{array}
  \right)
  }_{\mathrm{F}}\\
  \label{eq:phi_m_2}
  &= \phi_{m-1;2i,2j}(x,y) + \phi_{m-1;2i+1,2j}(x,y)
  + \phi_{m-1;2i,2j+1}(x,y) + \phi_{m-1;2i,2j+1}(x,y)
\end{align}
where $\braket{\mathsf{A}|\mathsf{B}}_{\mathrm{F}}$ denotes
the Frobenius inner product of matrices $\mathsf{A}$ and $\mathsf{B}$.
The scaling functions satisfy orthogonal property,
\begin{align}
  \label{eq:phi_orthogonality1}
  \braket{\phi_{m;i,j}(x,y)|\phi_{m;i',j'}(x,y)}
  &= \sum_{x,y}\phi_{m;i,j}(x,y),\phi_{m;i',j'}(x,y)\\
  \label{eq:phi_orthogonality2}
  &= 4^m\delta_{i,i'}\delta_{j,j'},
\end{align}
where the summation runs all the lattice points,
$x,y \in [0,2^n-1]$,
and $\delta_{i,j}$ is the Kronecker delta.
By virtue of the orthogonality, a scaling coefficient $c_{m;i,j}$
can be obtained by the following inner product,
\begin{equation}
  \label{eq:scaling_coefficient1}
  c_{m;i,j} = \braket{f(x,y)|\phi_{m;i,j}(x,y)}
\end{equation}
Using the recursion relation of $\phi_{m;i,j}(x,y)$, Eq.~(\ref{eq:phi_m_1}),
the above equation can be rewritten as
\begin{equation}
  \label{eq:scaling_coefficient2}
  c_{m;i,j} = \frac{1}{4}(c_{m-1;2i,2j} + c_{m-1;2i+1,2j} + c_{m-1;2i,2j+1} + c_{m-1;2i+1,2j+1}).
\end{equation}
Namely level-$m$ scaling coefficients are given by
the average of coefficients of level-$(m-1)$.
The 2D Haar wavelet functions are given by
\begin{align}
  \label{eq:wavelet_v}
  \psi^v_{m;i,j}(x,y)
  &= \frac{1}{4^m}
  \Braket{
  \left(
  \begin{array}{rr}
    1 & 1\\
    -1 & -1
  \end{array}
  \right)|
  \left(
  \begin{array}{ll}
    \phi_{m-1;2i,2j}(x,y) & \phi_{m-1;2i+1,2j}(x,y)\\
    \phi_{m-1;2i,2j+1}(x,y) & \phi_{m-1;2i,2j+1}(x,y)
  \end{array}
  \right)
  }_{\mathrm{F}},\\
  \label{eq:wavelet_h}
  \psi^h_{m;i,j}(x,y)
  &= \frac{1}{4^m}
  \Braket{
  \left(
  \begin{array}{rr}
    1 & -1\\
    1 & -1
  \end{array}
  \right)|
  \left(
  \begin{array}{ll}
    \phi_{m-1;2i,2j}(x,y) & \phi_{m-1;2i+1,2j}(x,y)\\
    \phi_{m-1;2i,2j+1}(x,y) & \phi_{m-1;2i,2j+1}(x,y)
  \end{array}
  \right)
  }_{\mathrm{F}},\\
  \label{eq:wavelet_d}
  \psi^d_{m;i,j}(x,y)
  &= \frac{1}{4^m}
  \Braket{
  \left(
  \begin{array}{rr}
    1 & -1\\
    -1 & 1
  \end{array}
  \right)|
  \left(
  \begin{array}{ll}
    \phi_{m-1;2i,2j}(x,y) & \phi_{m-1;2i+1,2j}(x,y)\\
    \phi_{m-1;2i,2j+1}(x,y) & \phi_{m-1;2i,2j+1}(x,y)
  \end{array}
  \right)
  }_{\mathrm{F}}.
\end{align}
These three wavelet functions are also orthogonal:
\begin{equation}
  \label{eq:psi_orthogonality}
  \braket{\psi^\alpha_{m;i,j}(x,y)|\psi^\beta_{m;i',j'}(x,y)}
  = \frac{1}{4^m}\delta_{\alpha,\beta}\delta_{i,i'}\delta_{j,j'},
  \quad(\alpha, \beta \in \{v, h, d\}).
\end{equation}
Wavelet coefficients $d^\alpha_{m;i,j}$ can be obtained by
\begin{align}
  \label{eq:wavelet_coefficient_v1}
  d^v_{m;i,j} &= \braket{f(x,y)|\psi^v_{m;i,j}(x,y)},\\
  \label{eq:wavelet_coefficient_h1}
  d^h_{m;i,j} &= \braket{f(x,y)|\psi^h_{m;i,j}(x,y)},\\
  \label{eq:wavelet_coefficient_d1}
  d^d_{m;i,j} &= \braket{f(x,y)|\psi^d_{m;i,j}(x,y)}.
\end{align}
Again, using recursive relation we obtain
\begin{align}
  \label{eq:wavelet_coefficient_v2}
  d^v_{m;i,j} &= \frac{1}{4}(c_{m-1;2i,2j} + c_{m-1;2i+1,2j} - c_{m-1;2i,2j+1} - c_{m-1;2i+1,2j+1}),\\
  \label{eq:wavelet_coefficient_h2}
  d^h_{m;i,j} &= \frac{1}{4}(c_{m-1;2i,2j} - c_{m-1;2i+1,2j} + c_{m-1;2i,2j+1} - c_{m-1;2i+1,2j+1}),\\
  \label{eq:wavelet_coefficient_d2}
  d^d_{m;i,j} &= \frac{1}{4}(c_{m-1;2i,2j} - c_{m-1;2i+1,2j} - c_{m-1;2i,2j+1} + c_{m-1;2i+1,2j+1}).
\end{align}
From these equations, it can be read that
level-$m$ wavelet coefficients preserve differences
between neighboring $f_{m-1}(x,y)$'s.
Indeed, level-$(m-1)$ scaling functions can be restored by
level-$m$ scaling and wavelet functions:
\begin{align}
  \phi_{m-1;2i,2j}(x,y) &= \frac{1}{4}\{\phi_{m;i,j}(x,y) + 4^m[\psi^v_{m;i,j}(x,y)
  + \psi^h_{m;i,j}(x,y) + \psi^d_{m;i,j}(x,y)]\},\\
  \phi_{m-1;2i+1,2j}(x,y) &= \frac{1}{4}\{\phi_{m;i,j}(x,y) + 4^m[\psi^v_{m;i,j}(x,y)
  - \psi^h_{m;i,j}(x,y) - \psi^d_{m;i,j}(x,y)]\},\\
  \phi_{m-1;2i,2j+1}(x,y) &= \frac{1}{4}\{\phi_{m;i,j}(x,y) + 4^m[-\psi^v_{m;i,j}(x,y)
  + \psi^h_{m;i,j}(x,y) - \psi^d_{m;i,j}(x,y)]\},\\
  \phi_{m-1;2i+1,2j+1}(x,y) &= \frac{1}{4}\{\phi_{m;i,j}(x,y) + 4^m[-\psi^v_{m;i,j}(x,y)
  - \psi^h_{m;i,j}(x,y) + \psi^d_{m;i,j}(x,y)]\}.
\end{align}
In a similar way, we obtain the following relations:
\begin{align}
  c_{m-1;2i,2j} &= c_{m;i,j} + d^v_{m;i,j} + d^h_{m;i,j} + d^d_{m;i,j},\\
  c_{m-1;2i+1,2j} &= c_{m;i,j} + d^v_{m;i,j} - d^h_{m;i,j} - d^d_{m;i,j},\\
  c_{m-1;2i,2j+1} &= c_{m;i,j} - d^v_{m;i,j} + d^h_{m;i,j} - d^d_{m;i,j},\\
  c_{m-1;2i+1,2j+1} &= c_{m;i,j} - d^v_{m;i,j} - d^h_{m;i,j} + d^d_{m;i,j}.
\end{align}
Namely the lost information caused by a wavelet transform
from level-$(m-1)$ to level-$m$ is properly preserved
in level-$m$ wavelet coefficients.
Therefore, I define a quantity which measures
an amount of emitted information entropy
at a level-$m$ wavelet transform as
\begin{equation}
  \mathcal{E}_m = 4^{m+1}\sum_{i,j}[(d^v_{m;i,j})^2 + (d^h_{m;i,j})^2 + (d^d_{m;i,j})^2].
\end{equation}
The normalized entropy emission is given by
\begin{equation}
  \epsilon_m = \frac{4^{m+1}}{3N_m}\sum_{i,j}[(d^v_{m;i,j})^2 + (d^h_{m;i,j})^2 + (d^d_{m;i,j})^2],
\end{equation}
where $N_m(=2^{n-m}\times 2^{n-m})$ is the number of coarse-grained lattice sites.
The number 3 in the denominator makes the normalized entropy emission
unity in the critical percolation problem.
In this paper, we applied the 2D Haar wavelet transform to
vertical and horizontal bonds, and contributions from the both kinds of bonds
are averaged.
The original function $f(x,y)$ is set as unity (zero)
if a bond is occupied (empty) at $(x,y)$.
Therefore the coarse-graining procedure for bonds is essentially the same
as in the demonstration in Sec.~\ref{sec:method}.
\end{widetext}

\bibliography{tomita}

\begin{thebibliography}{36}%
\makeatletter
\providecommand \@ifxundefined [1]{%
 \@ifx{#1\undefined}
}%
\providecommand \@ifnum [1]{%
 \ifnum #1\expandafter \@firstoftwo
 \else \expandafter \@secondoftwo
 \fi
}%
\providecommand \@ifx [1]{%
 \ifx #1\expandafter \@firstoftwo
 \else \expandafter \@secondoftwo
 \fi
}%
\providecommand \natexlab [1]{#1}%
\providecommand \enquote  [1]{``#1''}%
\providecommand \bibnamefont  [1]{#1}%
\providecommand \bibfnamefont [1]{#1}%
\providecommand \citenamefont [1]{#1}%
\providecommand \href@noop [0]{\@secondoftwo}%
\providecommand \href [0]{\begingroup \@sanitize@url \@href}%
\providecommand \@href[1]{\@@startlink{#1}\@@href}%
\providecommand \@@href[1]{\endgroup#1\@@endlink}%
\providecommand \@sanitize@url [0]{\catcode `\\12\catcode `\$12\catcode
  `\&12\catcode `\#12\catcode `\^12\catcode `\_12\catcode `\%12\relax}%
\providecommand \@@startlink[1]{}%
\providecommand \@@endlink[0]{}%
\providecommand \url  [0]{\begingroup\@sanitize@url \@url }%
\providecommand \@url [1]{\endgroup\@href {#1}{\urlprefix }}%
\providecommand \urlprefix  [0]{URL }%
\providecommand \Eprint [0]{\href }%
\providecommand \doibase [0]{http://dx.doi.org/}%
\providecommand \selectlanguage [0]{\@gobble}%
\providecommand \bibinfo  [0]{\@secondoftwo}%
\providecommand \bibfield  [0]{\@secondoftwo}%
\providecommand \translation [1]{[#1]}%
\providecommand \BibitemOpen [0]{}%
\providecommand \bibitemStop [0]{}%
\providecommand \bibitemNoStop [0]{.\EOS\space}%
\providecommand \EOS [0]{\spacefactor3000\relax}%
\providecommand \BibitemShut  [1]{\csname bibitem#1\endcsname}%
\let\auto@bib@innerbib\@empty
\bibitem [{\citenamefont {White}(1992)}]{White1992}%
  \BibitemOpen
  \bibfield  {author} {\bibinfo {author} {\bibfnamefont {S.~R.}\ \bibnamefont
  {White}},\ }\href {\doibase 10.1103/PhysRevLett.69.2863} {\bibfield
  {journal} {\bibinfo  {journal} {Phys. Rev. Lett.}\ }\textbf {\bibinfo
  {volume} {69}},\ \bibinfo {pages} {2863} (\bibinfo {year}
  {1992})}\BibitemShut {NoStop}%
\bibitem [{\citenamefont {White}(1993)}]{White1993}%
  \BibitemOpen
  \bibfield  {author} {\bibinfo {author} {\bibfnamefont {S.~R.}\ \bibnamefont
  {White}},\ }\href {\doibase 10.1103/PhysRevB.48.10345} {\bibfield  {journal}
  {\bibinfo  {journal} {Phys. Rev. B}\ }\textbf {\bibinfo {volume} {48}},\
  \bibinfo {pages} {10345} (\bibinfo {year} {1993})}\BibitemShut {NoStop}%
\bibitem [{\citenamefont {Nishino}\ and\ \citenamefont
  {Okunishi}(1996)}]{NishinoOkunishi1996}%
  \BibitemOpen
  \bibfield  {author} {\bibinfo {author} {\bibfnamefont {T.}~\bibnamefont
  {Nishino}}\ and\ \bibinfo {author} {\bibfnamefont {K.}~\bibnamefont
  {Okunishi}},\ }\href {\doibase 10.1143/JPSJ.65.891} {\bibfield  {journal}
  {\bibinfo  {journal} {J. Phys. Soc. Jpn.}\ }\textbf {\bibinfo {volume}
  {65}},\ \bibinfo {pages} {891} (\bibinfo {year} {1996})}\BibitemShut
  {NoStop}%
\bibitem [{\citenamefont {Nishino}\ and\ \citenamefont
  {Okunishi}(1997)}]{NishinoOkunishi1997}%
  \BibitemOpen
  \bibfield  {author} {\bibinfo {author} {\bibfnamefont {T.}~\bibnamefont
  {Nishino}}\ and\ \bibinfo {author} {\bibfnamefont {K.}~\bibnamefont
  {Okunishi}},\ }\href {\doibase 10.1143/JPSJ.66.3040} {\bibfield  {journal}
  {\bibinfo  {journal} {J. Phys. Soc. Jpn.}\ }\textbf {\bibinfo {volume}
  {66}},\ \bibinfo {pages} {3040} (\bibinfo {year} {1997})}\BibitemShut
  {NoStop}%
\bibitem [{\citenamefont {Levin}\ and\ \citenamefont
  {Nave}(2007)}]{LevinNave2007}%
  \BibitemOpen
  \bibfield  {author} {\bibinfo {author} {\bibfnamefont {M.}~\bibnamefont
  {Levin}}\ and\ \bibinfo {author} {\bibfnamefont {C.~P.}\ \bibnamefont
  {Nave}},\ }\href {\doibase 10.1103/PhysRevLett.99.120601} {\bibfield
  {journal} {\bibinfo  {journal} {Phys. Rev. Lett.}\ }\textbf {\bibinfo
  {volume} {99}},\ \bibinfo {pages} {120601} (\bibinfo {year}
  {2007})}\BibitemShut {NoStop}%
\bibitem [{\citenamefont {Evenbly}\ and\ \citenamefont
  {Vidal}(2009)}]{EvenblyVidal2009}%
  \BibitemOpen
  \bibfield  {author} {\bibinfo {author} {\bibfnamefont {G.}~\bibnamefont
  {Evenbly}}\ and\ \bibinfo {author} {\bibfnamefont {G.}~\bibnamefont
  {Vidal}},\ }\href {\doibase 10.1103/PhysRevB.79.144108} {\bibfield  {journal}
  {\bibinfo  {journal} {Phys. Rev. B}\ }\textbf {\bibinfo {volume} {79}},\
  \bibinfo {pages} {144108} (\bibinfo {year} {2009})}\BibitemShut {NoStop}%
\bibitem [{\citenamefont {Evenbly}\ and\ \citenamefont
  {Vidal}(2015)}]{EvenblyVidal2015}%
  \BibitemOpen
  \bibfield  {author} {\bibinfo {author} {\bibfnamefont {G.}~\bibnamefont
  {Evenbly}}\ and\ \bibinfo {author} {\bibfnamefont {G.}~\bibnamefont
  {Vidal}},\ }\href {\doibase 10.1103/PhysRevLett.115.180405} {\bibfield
  {journal} {\bibinfo  {journal} {Phys. Rev. Lett.}\ }\textbf {\bibinfo
  {volume} {115}},\ \bibinfo {pages} {180405} (\bibinfo {year}
  {2015})}\BibitemShut {NoStop}%
\bibitem [{\citenamefont {Ma}(1976)}]{Ma1976}%
  \BibitemOpen
  \bibfield  {author} {\bibinfo {author} {\bibfnamefont {S.-k.}\ \bibnamefont
  {Ma}},\ }\href {\doibase 10.1103/PhysRevLett.37.461} {\bibfield  {journal}
  {\bibinfo  {journal} {Phys. Rev. Lett.}\ }\textbf {\bibinfo {volume} {37}},\
  \bibinfo {pages} {461} (\bibinfo {year} {1976})}\BibitemShut {NoStop}%
\bibitem [{\citenamefont {Swendsen}(1979)}]{Swendsen1979}%
  \BibitemOpen
  \bibfield  {author} {\bibinfo {author} {\bibfnamefont {R.~H.}\ \bibnamefont
  {Swendsen}},\ }\href {\doibase 10.1103/PhysRevLett.42.859} {\bibfield
  {journal} {\bibinfo  {journal} {Phys. Rev. Lett.}\ }\textbf {\bibinfo
  {volume} {42}},\ \bibinfo {pages} {859} (\bibinfo {year} {1979})}\BibitemShut
  {NoStop}%
\bibitem [{\citenamefont {Sugiura}\ and\ \citenamefont
  {Shimizu}(2013)}]{SugiuraShimizu2013}%
  \BibitemOpen
  \bibfield  {author} {\bibinfo {author} {\bibfnamefont {S.}~\bibnamefont
  {Sugiura}}\ and\ \bibinfo {author} {\bibfnamefont {A.}~\bibnamefont
  {Shimizu}},\ }\href {\doibase 10.1103/PhysRevLett.111.010401} {\bibfield
  {journal} {\bibinfo  {journal} {Phys. Rev. Lett.}\ }\textbf {\bibinfo
  {volume} {111}},\ \bibinfo {pages} {010401} (\bibinfo {year}
  {2013})}\BibitemShut {NoStop}%
\bibitem [{\citenamefont {Tasaki}(2016)}]{Tasaki2016}%
  \BibitemOpen
  \bibfield  {author} {\bibinfo {author} {\bibfnamefont {H.}~\bibnamefont
  {Tasaki}},\ }\href@noop {} {\bibfield  {journal} {\bibinfo  {journal}
  {Journal of Statistical Physics}\ }\textbf {\bibinfo {volume} {163}},\
  \bibinfo {pages} {937} (\bibinfo {year} {2016})}\BibitemShut {NoStop}%
\bibitem [{\citenamefont {Ueda}\ \emph {et~al.}(2005)\citenamefont {Ueda},
  \citenamefont {Otani}, \citenamefont {Nishino}, \citenamefont {Gendiar},\
  and\ \citenamefont {Nishino}}]{Ueda2005}%
  \BibitemOpen
  \bibfield  {author} {\bibinfo {author} {\bibfnamefont {K.}~\bibnamefont
  {Ueda}}, \bibinfo {author} {\bibfnamefont {R.}~\bibnamefont {Otani}},
  \bibinfo {author} {\bibfnamefont {Y.}~\bibnamefont {Nishino}}, \bibinfo
  {author} {\bibfnamefont {A.}~\bibnamefont {Gendiar}}, \ and\ \bibinfo
  {author} {\bibfnamefont {T.}~\bibnamefont {Nishino}},\ }\href@noop {}
  {\bibfield  {journal} {\bibinfo  {journal} {J. Phys. Soc. Jpn. Suppl.}\
  }\textbf {\bibinfo {volume} {74}},\ \bibinfo {pages} {111} (\bibinfo {year}
  {2005})}\BibitemShut {NoStop}%
\bibitem [{\citenamefont {Matsueda}(2012)}]{Matsueda2012}%
  \BibitemOpen
  \bibfield  {author} {\bibinfo {author} {\bibfnamefont {H.}~\bibnamefont
  {Matsueda}},\ }\href {\doibase 10.1103/PhysRevE.85.031101} {\bibfield
  {journal} {\bibinfo  {journal} {Phys. Rev. E}\ }\textbf {\bibinfo {volume}
  {85}},\ \bibinfo {pages} {031101} (\bibinfo {year} {2012})}\BibitemShut
  {NoStop}%
\bibitem [{\citenamefont {Imura}\ \emph {et~al.}(2014)\citenamefont {Imura},
  \citenamefont {Okubo}, \citenamefont {Morita},\ and\ \citenamefont
  {Okunishi}}]{Imura2014}%
  \BibitemOpen
  \bibfield  {author} {\bibinfo {author} {\bibfnamefont {Y.}~\bibnamefont
  {Imura}}, \bibinfo {author} {\bibfnamefont {T.}~\bibnamefont {Okubo}},
  \bibinfo {author} {\bibfnamefont {S.}~\bibnamefont {Morita}}, \ and\ \bibinfo
  {author} {\bibfnamefont {K.}~\bibnamefont {Okunishi}},\ }\href {\doibase
  10.7566/JPSJ.83.114002} {\bibfield  {journal} {\bibinfo  {journal} {J. Phys.
  Soc. Jpn.}\ }\textbf {\bibinfo {volume} {83}},\ \bibinfo {pages} {114002}
  (\bibinfo {year} {2014})}\BibitemShut {NoStop}%
\bibitem [{\citenamefont {Lee}\ \emph {et~al.}(2015)\citenamefont {Lee},
  \citenamefont {Yamada}, \citenamefont {Kumamoto},\ and\ \citenamefont
  {Matsueda}}]{Lee2015}%
  \BibitemOpen
  \bibfield  {author} {\bibinfo {author} {\bibfnamefont {C.~H.}\ \bibnamefont
  {Lee}}, \bibinfo {author} {\bibfnamefont {Y.}~\bibnamefont {Yamada}},
  \bibinfo {author} {\bibfnamefont {T.}~\bibnamefont {Kumamoto}}, \ and\
  \bibinfo {author} {\bibfnamefont {H.}~\bibnamefont {Matsueda}},\ }\href
  {\doibase 10.7566/JPSJ.84.013001} {\bibfield  {journal} {\bibinfo  {journal}
  {J. Phys. Soc. Jpn.}\ }\textbf {\bibinfo {volume} {84}},\ \bibinfo {pages}
  {013001} (\bibinfo {year} {2015})}\BibitemShut {NoStop}%
\bibitem [{\citenamefont {Matsueda}\ and\ \citenamefont
  {Ozaki}(2015)}]{MatsuedaOzaki2015}%
  \BibitemOpen
  \bibfield  {author} {\bibinfo {author} {\bibfnamefont {H.}~\bibnamefont
  {Matsueda}}\ and\ \bibinfo {author} {\bibfnamefont {D.}~\bibnamefont
  {Ozaki}},\ }\href {\doibase 10.1103/PhysRevE.92.042167} {\bibfield  {journal}
  {\bibinfo  {journal} {Phys. Rev. E}\ }\textbf {\bibinfo {volume} {92}},\
  \bibinfo {pages} {042167} (\bibinfo {year} {2015})}\BibitemShut {NoStop}%
\bibitem [{\citenamefont {Lee}\ \emph {et~al.}(2016)\citenamefont {Lee},
  \citenamefont {Ozaki},\ and\ \citenamefont {Matsueda}}]{Lee2016}%
  \BibitemOpen
  \bibfield  {author} {\bibinfo {author} {\bibfnamefont {C.~H.}\ \bibnamefont
  {Lee}}, \bibinfo {author} {\bibfnamefont {D.}~\bibnamefont {Ozaki}}, \ and\
  \bibinfo {author} {\bibfnamefont {H.}~\bibnamefont {Matsueda}},\ }\href
  {\doibase 10.1103/PhysRevE.94.062144} {\bibfield  {journal} {\bibinfo
  {journal} {Phys. Rev. E}\ }\textbf {\bibinfo {volume} {94}},\ \bibinfo
  {pages} {062144} (\bibinfo {year} {2016})}\BibitemShut {NoStop}%
\bibitem [{\citenamefont {Daubechies}(1992)}]{Daubechies}%
  \BibitemOpen
  \bibfield  {author} {\bibinfo {author} {\bibfnamefont {I.}~\bibnamefont
  {Daubechies}},\ }\href@noop {} {\emph {\bibinfo {title} {Ten Lectures on
  Wavelets}}}\ (\bibinfo  {publisher} {Society for Industrial and Applied
  Mathematics},\ \bibinfo {year} {1992})\BibitemShut {NoStop}%
\bibitem [{\citenamefont {Kasteleyn}\ and\ \citenamefont
  {Fortuin}(1969)}]{KasteleynFortuin1969}%
  \BibitemOpen
  \bibfield  {author} {\bibinfo {author} {\bibfnamefont {P.}~\bibnamefont
  {Kasteleyn}}\ and\ \bibinfo {author} {\bibfnamefont {C.}~\bibnamefont
  {Fortuin}},\ }\href@noop {} {\bibfield  {journal} {\bibinfo  {journal} {J.
  Phys. Soc. Jpn. Suppl.}\ }\textbf {\bibinfo {volume} {26}},\ \bibinfo {pages}
  {11} (\bibinfo {year} {1969})}\BibitemShut {NoStop}%
\bibitem [{\citenamefont {Fortuin}\ and\ \citenamefont
  {Kasteleyn}(1972)}]{FortuinKasteleyn1972}%
  \BibitemOpen
  \bibfield  {author} {\bibinfo {author} {\bibfnamefont {C.}~\bibnamefont
  {Fortuin}}\ and\ \bibinfo {author} {\bibfnamefont {P.}~\bibnamefont
  {Kasteleyn}},\ }\href {\doibase https://doi.org/10.1016/0031-8914(72)90045-6}
  {\bibfield  {journal} {\bibinfo  {journal} {Physica}\ }\textbf {\bibinfo
  {volume} {57}},\ \bibinfo {pages} {536 } (\bibinfo {year}
  {1972})}\BibitemShut {NoStop}%
\bibitem [{\citenamefont {Holzhey}\ \emph {et~al.}(1994)\citenamefont
  {Holzhey}, \citenamefont {Larsen},\ and\ \citenamefont
  {Wilczek}}]{Holzhey1994}%
  \BibitemOpen
  \bibfield  {author} {\bibinfo {author} {\bibfnamefont {C.}~\bibnamefont
  {Holzhey}}, \bibinfo {author} {\bibfnamefont {F.}~\bibnamefont {Larsen}}, \
  and\ \bibinfo {author} {\bibfnamefont {F.}~\bibnamefont {Wilczek}},\
  }\href@noop {} {\bibfield  {journal} {\bibinfo  {journal} {Nuclear Physics
  B}\ }\textbf {\bibinfo {volume} {424}},\ \bibinfo {pages} {443 } (\bibinfo
  {year} {1994})}\BibitemShut {NoStop}%
\bibitem [{\citenamefont {Calabrese}\ and\ \citenamefont
  {Cardy}(2004)}]{CalabreseCardy2004}%
  \BibitemOpen
  \bibfield  {author} {\bibinfo {author} {\bibfnamefont {P.}~\bibnamefont
  {Calabrese}}\ and\ \bibinfo {author} {\bibfnamefont {J.}~\bibnamefont
  {Cardy}},\ }\href@noop {} {\bibfield  {journal} {\bibinfo  {journal} {J.
  Stat. Mech.}\ }\textbf {\bibinfo {volume} {2004}},\ \bibinfo {pages} {P06002}
  (\bibinfo {year} {2004})}\BibitemShut {NoStop}%
\bibitem [{\citenamefont {Potts}(1952)}]{Potts1952}%
  \BibitemOpen
  \bibfield  {author} {\bibinfo {author} {\bibfnamefont {R.~B.}\ \bibnamefont
  {Potts}},\ }\href {\doibase 10.1017/S0305004100027419} {\bibfield  {journal}
  {\bibinfo  {journal} {Mathematical Proceedings of the Cambridge Philosophical
  Society}\ }\textbf {\bibinfo {volume} {48}},\ \bibinfo {pages} {106–109}
  (\bibinfo {year} {1952})}\BibitemShut {NoStop}%
\bibitem [{\citenamefont {Kihara}\ \emph {et~al.}(1954)\citenamefont {Kihara},
  \citenamefont {Midzuno},\ and\ \citenamefont {Shizume}}]{Kihara1954}%
  \BibitemOpen
  \bibfield  {author} {\bibinfo {author} {\bibfnamefont {T.}~\bibnamefont
  {Kihara}}, \bibinfo {author} {\bibfnamefont {Y.}~\bibnamefont {Midzuno}}, \
  and\ \bibinfo {author} {\bibfnamefont {T.}~\bibnamefont {Shizume}},\ }\href
  {\doibase 10.1143/JPSJ.9.681} {\bibfield  {journal} {\bibinfo  {journal} {J.
  Phys. Soc. Jpn.}\ }\textbf {\bibinfo {volume} {9}},\ \bibinfo {pages} {681}
  (\bibinfo {year} {1954})}\BibitemShut {NoStop}%
\bibitem [{\citenamefont {Wu}(1982)}]{Wu1982}%
  \BibitemOpen
  \bibfield  {author} {\bibinfo {author} {\bibfnamefont {F.~Y.}\ \bibnamefont
  {Wu}},\ }\href {\doibase 10.1103/RevModPhys.54.235} {\bibfield  {journal}
  {\bibinfo  {journal} {Rev. Mod. Phys.}\ }\textbf {\bibinfo {volume} {54}},\
  \bibinfo {pages} {235} (\bibinfo {year} {1982})}\BibitemShut {NoStop}%
\bibitem [{\citenamefont {Coniglio}\ and\ \citenamefont
  {Klein}(1980)}]{ConiglioKlein1980}%
  \BibitemOpen
  \bibfield  {author} {\bibinfo {author} {\bibfnamefont {A.}~\bibnamefont
  {Coniglio}}\ and\ \bibinfo {author} {\bibfnamefont {W.}~\bibnamefont
  {Klein}},\ }\href {http://stacks.iop.org/0305-4470/13/i=8/a=025} {\bibfield
  {journal} {\bibinfo  {journal} {Journal of Physics A: Mathematical and
  General}\ }\textbf {\bibinfo {volume} {13}},\ \bibinfo {pages} {2775}
  (\bibinfo {year} {1980})}\BibitemShut {NoStop}%
\bibitem [{\citenamefont {Hu}(1984)}]{Hu1984a}%
  \BibitemOpen
  \bibfield  {author} {\bibinfo {author} {\bibfnamefont {C.-K.}\ \bibnamefont
  {Hu}},\ }\href {\doibase 10.1103/PhysRevB.29.5103} {\bibfield  {journal}
  {\bibinfo  {journal} {Phys. Rev. B}\ }\textbf {\bibinfo {volume} {29}},\
  \bibinfo {pages} {5103} (\bibinfo {year} {1984})}\BibitemShut {NoStop}%
\bibitem [{\citenamefont {Swendsen}\ and\ \citenamefont
  {Wang}(1987)}]{SwendsenWang1987}%
  \BibitemOpen
  \bibfield  {author} {\bibinfo {author} {\bibfnamefont {R.~H.}\ \bibnamefont
  {Swendsen}}\ and\ \bibinfo {author} {\bibfnamefont {J.-S.}\ \bibnamefont
  {Wang}},\ }\href {\doibase 10.1103/PhysRevLett.58.86} {\bibfield  {journal}
  {\bibinfo  {journal} {Phys. Rev. Lett.}\ }\textbf {\bibinfo {volume} {58}},\
  \bibinfo {pages} {86} (\bibinfo {year} {1987})}\BibitemShut {NoStop}%
\bibitem [{\citenamefont {Komura}\ and\ \citenamefont
  {Okabe}(2012)}]{KomuraOkabe2012}%
  \BibitemOpen
  \bibfield  {author} {\bibinfo {author} {\bibfnamefont {Y.}~\bibnamefont
  {Komura}}\ and\ \bibinfo {author} {\bibfnamefont {Y.}~\bibnamefont {Okabe}},\
  }\href {\doibase https://doi.org/10.1016/j.cpc.2012.01.017} {\bibfield
  {journal} {\bibinfo  {journal} {Computer Physics Communications}\ }\textbf
  {\bibinfo {volume} {183}},\ \bibinfo {pages} {1155 } (\bibinfo {year}
  {2012})}\BibitemShut {NoStop}%
\bibitem [{\citenamefont {Komura}(2015)}]{Komura2015}%
  \BibitemOpen
  \bibfield  {author} {\bibinfo {author} {\bibfnamefont {Y.}~\bibnamefont
  {Komura}},\ }\href {\doibase https://doi.org/10.1016/j.cpc.2015.04.015}
  {\bibfield  {journal} {\bibinfo  {journal} {Computer Physics Communications}\
  }\textbf {\bibinfo {volume} {194}},\ \bibinfo {pages} {54 } (\bibinfo {year}
  {2015})}\BibitemShut {NoStop}%
\bibitem [{\citenamefont {Duplantier}(2000)}]{Duplantier2000}%
  \BibitemOpen
  \bibfield  {author} {\bibinfo {author} {\bibfnamefont {B.}~\bibnamefont
  {Duplantier}},\ }\href {\doibase 10.1103/PhysRevLett.84.1363} {\bibfield
  {journal} {\bibinfo  {journal} {Phys. Rev. Lett.}\ }\textbf {\bibinfo
  {volume} {84}},\ \bibinfo {pages} {1363} (\bibinfo {year}
  {2000})}\BibitemShut {NoStop}%
\bibitem [{\citenamefont {Tomita}\ and\ \citenamefont
  {Okabe}(2001)}]{TomitaOkabe2001}%
  \BibitemOpen
  \bibfield  {author} {\bibinfo {author} {\bibfnamefont {Y.}~\bibnamefont
  {Tomita}}\ and\ \bibinfo {author} {\bibfnamefont {Y.}~\bibnamefont {Okabe}},\
  }\href {\doibase 10.1103/PhysRevLett.86.572} {\bibfield  {journal} {\bibinfo
  {journal} {Phys. Rev. Lett.}\ }\textbf {\bibinfo {volume} {86}},\ \bibinfo
  {pages} {572} (\bibinfo {year} {2001})}\BibitemShut {NoStop}%
\bibitem [{\citenamefont {Tomita}\ \emph {et~al.}(1999)\citenamefont {Tomita},
  \citenamefont {Okabe},\ and\ \citenamefont {Hu}}]{TomitaOkabeHu1999}%
  \BibitemOpen
  \bibfield  {author} {\bibinfo {author} {\bibfnamefont {Y.}~\bibnamefont
  {Tomita}}, \bibinfo {author} {\bibfnamefont {Y.}~\bibnamefont {Okabe}}, \
  and\ \bibinfo {author} {\bibfnamefont {C.-K.}\ \bibnamefont {Hu}},\ }\href
  {\doibase 10.1103/PhysRevE.60.2716} {\bibfield  {journal} {\bibinfo
  {journal} {Phys. Rev. E}\ }\textbf {\bibinfo {volume} {60}},\ \bibinfo
  {pages} {2716} (\bibinfo {year} {1999})}\BibitemShut {NoStop}%
\bibitem [{\citenamefont {Evertz}\ \emph {et~al.}(1993)\citenamefont {Evertz},
  \citenamefont {Lana},\ and\ \citenamefont {Marcu}}]{Evertz1993}%
  \BibitemOpen
  \bibfield  {author} {\bibinfo {author} {\bibfnamefont {H.~G.}\ \bibnamefont
  {Evertz}}, \bibinfo {author} {\bibfnamefont {G.}~\bibnamefont {Lana}}, \ and\
  \bibinfo {author} {\bibfnamefont {M.}~\bibnamefont {Marcu}},\ }\href
  {\doibase 10.1103/PhysRevLett.70.875} {\bibfield  {journal} {\bibinfo
  {journal} {Phys. Rev. Lett.}\ }\textbf {\bibinfo {volume} {70}},\ \bibinfo
  {pages} {875} (\bibinfo {year} {1993})}\BibitemShut {NoStop}%
\bibitem [{\citenamefont {Kawashima}\ and\ \citenamefont
  {Harada}(2004)}]{KawashimaHarada2004}%
  \BibitemOpen
  \bibfield  {author} {\bibinfo {author} {\bibfnamefont {N.}~\bibnamefont
  {Kawashima}}\ and\ \bibinfo {author} {\bibfnamefont {K.}~\bibnamefont
  {Harada}},\ }\href {\doibase 10.1143/JPSJ.73.1379} {\bibfield  {journal}
  {\bibinfo  {journal} {J. Phys. Soc. Jpn.}\ }\textbf {\bibinfo {volume}
  {73}},\ \bibinfo {pages} {1379} (\bibinfo {year} {2004})}\BibitemShut
  {NoStop}%
\bibitem [{\citenamefont {Tomita}\ and\ \citenamefont
  {Okabe}(2002)}]{TomitaOkabe2002}%
  \BibitemOpen
  \bibfield  {author} {\bibinfo {author} {\bibfnamefont {Y.}~\bibnamefont
  {Tomita}}\ and\ \bibinfo {author} {\bibfnamefont {Y.}~\bibnamefont {Okabe}},\
  }\href {\doibase 10.1103/PhysRevB.66.180401} {\bibfield  {journal} {\bibinfo
  {journal} {Phys. Rev. B}\ }\textbf {\bibinfo {volume} {66}},\ \bibinfo
  {pages} {180401} (\bibinfo {year} {2002})}\BibitemShut {NoStop}%
\end{thebibliography}%

\end{document}